\documentclass[prl,twocolumn,showpacs,showkeywords,preprintnumbers,amsmath,amssymb]{revtex4}
\usepackage{epsf}
\usepackage{amsfonts}
\usepackage{amssymb}
\usepackage{amsthm}
\usepackage{graphicx}
\newcommand{\cf}[1]{\langle #1 \rangle}                      
\newcommand{\bra}[1]{\langle #1 \!\mid\!}                    
\newcommand{\ket}[1]{\!\mid\! #1 \rangle}                    

\newcommand{\del}{\partial}
\newcommand{\bml}{\begin{mathletters}}
\newcommand{\eml}{\end{mathletters} \hspace{-5pt}}

\begin{document}

\title{Single-Site Entanglement of Fermions at a Quantum Phase Transition}

\author{Daniel Larsson$^{1,2}$ and Henrik Johannesson$^2$}
\affiliation{$\mbox{}^1$Fachbereich Physik, Philipps Universit\"at Marburg, D-35032 Marburg, Germany}
\affiliation{$\mbox{}^2$Department of Physics, G\"oteborg University, SE-412 96 G\"oteborg, Sweden}


\begin{abstract}

We show that the single-site entanglement of a generic spin-1/2 fermionic lattice system 
can be used as a reliable marker of a finite-order quantum phase transition, given certain provisos.
We discuss the information contained in the single-site entanglement measure, and provide illustrations from the
Mott-Hubbard metal-insulator transitions of the one-dimensional (1D) Hubbard model, and the (1D) Hubbard model with 
long-range hopping.

\end{abstract}



\maketitle

\newpage


{\bf Introduction.} \ The study of entanglement properties of many-particle systems has become a 
subject of intense interest.
Much of the motivation comes from quantum information theory where entanglement is made
the key physical resource for a variety of information processing tasks \cite{NielsenChuang}. 
In recent work it has been suggested that this resource may be efficiently extracted from
a solid, or from some other many-particle system, by scattering particles off the system \cite{DeChiara}. 
Thermodynamic properties of solids have also been shown to be
crucially influenced by entanglement properties of their microscopic degrees of freedom \cite{Ghosh}.
Moreover, a rapidly growing body of results \cite{Osterloh, Osborne} suggests that a properly chosen 
measure of entanglement may serve as a precise and convenient marker of a (zero-temperature) 
{\em quantum phase transition} (QPT) in a many-particle system \cite{Osterloh,Osborne}.
For spin-1/2 systems (lattices of localized coupled qubits) a discontinuity (divergence) in
the (derivative of the) ground state concurrence has been shown to be associated
with a first (second) order QPT
\cite{Wu} (where {\em concurrence} \cite{Wootters} measures the entanglement of two 
qubits selected at neighboring sites). For itinerant particles the picture is less clear, as the
results here appear to depend on the choice of model or on the perturbation driving the
transition. A case in point
is the {\em single-site entanglement} of the one-dimensional (1D) Hubbard model.
This measure, which is given by the von Neumann entropy at a single lattice site \cite{NielsenChuang},  
reaches a maximum at a metal-insulator transition driven by a change of
the on-site interaction \cite{Gu}. In contrast, the single-site entanglement
diverges when one drives the transition by tuning the chemical potential \cite{Larsson}. 

One should here realize that an onset of non-analyticity in a local entanglement measure \cite{footnote1}
is indeed expected at a QPT. By definition, a QPT is a point of
non-analyticity in the ground state energy of a quantum system (caused by a level crossing,
or, an avoided level crossing in the thermodynamic limit) \cite{Sachdev}.
Given that the elements of the reduced density matrix $-$ upon which any local entanglement measure is
built $-$ are linked to the ground state energy, the defining
non-analyticity of a QPT will infect also the local entanglement
measure (of which {\em single-site entanglement} \cite{NielsenChuang}, {\em concurrence} \cite{Wootters}, 
and {\em negativity} \cite{Vidal}
are some of the most commonly used). The recent proof that {\em any} entanglement measure can be expanded
as a unique functional of the first derivatives of the ground state energy (with respect to the parameters that
control the QPT) puts this intuition on firm ground \cite{Wu2}. 

The connection between entanglement and QPTs can also be cast in the language of statistical mechanics,
as pointed out recently by Campos Venuti {\em et al.} \cite{CamposVenuti}.
As an example, consider the Hamiltonian density ${\cal H}(g)$ of a system that undergoes
a continuous second-order QPT when changing a parameter $g$: ${\cal H}(g) = {\cal H}_0 + g\Lambda$. Differentiating
the energy density $\cf{\psi_0 | {\cal H}(g) | \psi_0}$ of the ground state $\ket{\psi_0}$ with respect to $g$,
its singular part ${\cal O}_g \sim [ \cf{\psi_0 | \Lambda | \psi_0} - \mbox{regular terms} ]$ will
behave as ${\cal O}_g \sim \mbox{sgn}(g-g_c)| g-g_c |^{\rho}$ as $g$ approaches $g_c$,
implying a divergence of $\del {\cal O}_g/\del g \sim
| g-g_c |^{\rho-1}$ at criticality.
The singular term ${\cal O}_g$ enters every reduced density matrix that
contains a site where the operator $\Lambda$ is defined, and it follows that any entanglement measure
constructed from such a density matrix exhibits a singularity with an exponent related to $\rho$ (barring
accidental cancellations).

Having established this linkage, one may ask how it can be exploited for a specific problem.
For example, in the case of a continuous second (or higher) order QPT, is it possible to ''read off'' the
critical exponent $\rho$ from the singularity of the entanglement measure? Conversely, is the information
provided by the singular behavior of a local entanglement measure already contained in the scaling of observables
$-$ as predicted within the usual statistical mechanics framework? 

In this article we address these questions by studying the 
single-site entanglement of a generic fermionic lattice system. We do so by 
constructing and analyzing its explicit representation using the 
Hellman-Feynman theorem.  We find that the single-site entanglement measure
can be used as reliable marker of a finite-order QPT (given certain provisos) and that it contains
unique and useful information about the transition. The questions raised above will both turn out to have 
negative answers. As illustrations we use our
construction to obtain the single-site entanglement at the Mott-Hubbard metal-insulator transitions of
the 1D Hubbard model \cite{LiebWu}, and the 1D Hubbard model with long-range hopping
\cite{GebhardRuckenstein}, exploiting exact results for the ground state properties of these models. We stress that
our analysis can be easily adapted so as to apply to a system of localized spins, with no change in the
general results. Specifically, the questions raised above are answered in the negative also for coupled qubit 
(spin-1/2) systems.
Our reason for focusing on fermionic systems is simply that these are less well understood.
With our contribution we hope to dispel some of the perceived difficulties attached to their treatment.

{\bf Single-site entanglement and QPTs.} \ Let us first recall that the concept of quantum entanglement
of indistinguishable fermions [bosons] suffers from a certain ambiguity since the accessible state space
contains only antisymmetrized [symmetrized] states and hence lacks a direct product structure. The simplest
way around this problem is to use an occupation number representation \cite{Zanardi}. For spin-1/2
fermions one thus takes 
$\ket{n}_j = \,\ket{0}_j, \,$ \nolinebreak[0]{$\ket{\,\uparrow}_j$}$, \,\ket{\,\downarrow}_j$, and
$\ket{\,\uparrow \downarrow}_j$ as local basis states, with $j\!=\!1,2,...,L$ indexing the corresponding
lattice sites. In this way the product structure of the state space is manifestly recovered, with the
representation spanned by the $4^L$ basis states $\ket{n}_1 \otimes \ket{n}_2 \otimes
 ... \otimes \ket{n}_L $. One may now proceed as usual and partition the system into two parts A and B,
with the entanglement (von Neumann) entropy ${\cal E}$ of a pure state $\ket{\psi}$ defined by \cite{NielsenChuang}
\begin{equation} \label{vonNeumann}
{\cal E} = - \mbox{Tr}(\rho_A \mbox{log}_2 \rho_A).
\end{equation}
The reduced density matrix $\rho_A$ is calculated from the full density matrix $\rho =
\,\ket{\psi} \bra{\psi}$ by taking the trace over the local states belonging to B: $\rho_A = \mbox{Tr}_B(\rho)$.
By choosing A as a single site (assuming translational invariance) with B the rest of the
system, one obtains the {\em single-site entanglement}. 
One should note that in the occupation number representation the subsystems A and B correspond
to {\em fermionic modes} (empty sites, singly occupied sites with spin up or down, doubly occupied sites)
and not to particles. In this sense the notion of fermionic (and similarly, bosonic) entanglement is
different from the text book example with spatially separated particles. 

Given the occupation number representation it is straightforward to verify that the reduced 
ground state density matrix $\rho_j$ for a single site $j$ is diagonal, provided that the 
ground state $\ket{\psi_0}$ is a superposition of basis states with the same number of
particles and the same total spin.   
Introducing the ground state expectation
values for a single site to be doubly occupied $(w_2)$, singly occupied by a fermion with spin-up [spin-down],
$(w_{\uparrow [\downarrow]})$, or empty $(w_0)$, and assuming that the system is translationally invariant,
we write:
\begin{eqnarray}  \label{OccupationParameters}
w_2 &=& \cf{\psi_0| \hat{n}_{j \uparrow} \hat{n}_{j \downarrow} | \psi_0} \nonumber \\
w_{\uparrow} &=& \cf{\psi_0 | \hat{n}_{j \uparrow} | \psi_0} - w_2 = \frac{n}{2} + m - w_2 \nonumber \\
w_{\downarrow} &=& \cf{\psi_0 | \hat{n}_{j \downarrow} | \psi_0} - w_2 = \frac{n}{2} - m - w_2 \nonumber \\
w_0 &=& 1-n+w_2
\end{eqnarray}
where in Eq. (\ref{OccupationParameters})
$\hat{n}_{j \sigma} = \hat{c}^{\dagger}_{j \sigma} \hat{c}_{j \sigma}$ is the number operator that samples site $j$
for a fermion of spin $\sigma = \uparrow, \downarrow$, $n = \cf{\psi_0 | \hat{n}_{j \uparrow} +
\hat{n}_{j \downarrow} | \psi_0}$ is the average single site occupation in the ground state, and
$m = (1/2)\cf{\psi_0 | \hat{n}_{j \uparrow} - \hat{n}_{j \downarrow} | \psi_0}$ is the
ground state magnetization per site.
It follows that
\begin{equation} \label{ReducedDensityMatrix}
\rho_j = \sum_{\alpha=0, \uparrow, \downarrow} w_{\alpha}\ket{\alpha}_j\bra{\alpha}_j 
+ w_2 \ket{\,\uparrow \downarrow}_j\bra{\uparrow \downarrow\,}_j.
\end{equation}
Combining Eqs. (\ref{vonNeumann}), (\ref{OccupationParameters}), and (\ref{ReducedDensityMatrix}) 
the single-site entanglement takes the form
\begin{eqnarray} \label{SingleSiteEntanglement}
{\cal E} & = & -\left(\frac{n}{2}+m-w_2\right)\log_2\left(\frac{n}{2}+m-w_2\right) \nonumber \\
         & - & \left(\frac{n}{2}-m-w_2\right)\log_2\left(\frac{n}{2}-m-w_2\right) - w_2\log_2w_2 \nonumber \\
         & - & \left(1-n+w_2\right)\log_2\left(1-n+w_2\right). 
\end{eqnarray}
Let us now consider a fermion system with Hamiltonian density ${\cal H}(g) = {\cal H}_0 + g\Lambda$ that
exhibits a QPT for some value $g_c$ of $g$ (with $\Lambda$ the conjugate operator, and with all other 
control parameters
kept fixed and absorbed as part of ${\cal H}_0$). By definition, a QPT of $k^{\mbox{\small th}}$ order implies a 
divergence or a discontinuity in
the $k^{\mbox{\small th}}$ derivative $\partial^k e_0/\partial g^k$ of the ground state energy density 
$e_0 = \cf{\psi_0 | {\cal H}(g) | \psi_0}$, with all derivatives of order $\!<\!k$ being finite and continuous. 
Defining ${\cal O}_g \equiv [\cf{\psi_0 | \Lambda | \psi_0} - \mbox{regular terms}]$ 
(equal to [$\partial e_0/\partial g$ - regular terms] by the Hellman-Feynman theorem), 
it follows that $\partial^{k-1}{\cal O}_g/\partial g^{k-1}$ has a divergence or a discontinuity at $g=g_c$.
With these preliminaries we can now prove the following \\

{\em Proposition} \\
Consider a spin-1/2 translationally invariant fermionic system with a Hamiltonian density ${\cal H}(g) = {\cal H}_0 +
g\Lambda$ that conserves particle number and total spin, and where 
${\cal O}_g \equiv [\cf{\psi_0 | \Lambda | \psi_0} - \mbox{regular terms}]$ is a linear combination of $m, n$ and/or
$w_2$. 
It follows that a divergence or a discontinuity in the $(k-1)^{\mbox{\small st}}$ derivative of the single-site entanglement
with respect to $g$
(with all derivatives of order $\!<k-\!1\!$ being finite and continuous) signals that the system undergoes
a $k^{\mbox{\small th}}$ order QPT. 

{\em Proof} \\
The proof is elementary. Repeated differentiation of Eq. (\ref{SingleSiteEntanglement}) yields
\begin{widetext}
\begin{multline} \label{n:thOrder}
\frac{ \partial^{k-1} {\cal E} }{ \partial g^{k-1} }=-\left(\frac{ \partial^{k-1} }
{ \partial g^{k-1} }[\frac{n}{2}+ m
-w_2]\right)\log_2\left(\frac{n}{2}+m-w_2\right)  
-\left(\frac{ \partial^{k-1} }{ \partial g^{k-1} }[\frac{n}{2}- m
-w_2]\right)\log_2\left(\frac{n}{2}-m-w_2\right) \\ 
-\frac{ \partial^{k-1} w_2 }{ \partial g^{k-1} }\log_2\left(w_2\right) 
+\left(\frac{ \partial^{k-1} }{ \partial g^{k-1} } 
[n-w_2]\right)\log_2\left(1-n+w_2\right) 
+\mbox{terms containing lower-order derivatives}.
\end{multline}
\end{widetext}
By assumption all derivatives with respect to $g$ of order $\!<k-\!1\!$ are finite and continuous. Any singularity in
$\partial^{k-1} 
{\cal E}/\partial g^{k-1}$ must hence reside in terms containing derivatives of order $k-1$.
Since ${\cal O}_g$ is a linear
combination of $m, n$ and $w_2$, the proposition follows. $\blacksquare$ \\

Several comments are in order. 
First note that the constraint that ${\cal O}_g$ should be some linear combination of $m, n$ and/or $w_2$ is
much less restrictive than may first appear to be the case. In fact, for a generic fermionic QPT caused by a 
change of an interaction or an external perturbation that
couples
only to single sites, ${\cal O}_g$ is {\em identical} to $w_2$
(with the transition driven by an on-site fermion-fermion interaction, $g\equiv u$), $m$ (with the
transition driven by a magnetic field, $g\equiv h$), or $n$
(with the transition driven by a chemical potential, $g\equiv \mu$). 
One may think that the tight link between the scaling of $\partial^{k-1} {\cal E}/
\partial g^{k-1}$ and that of $\partial^{k-1} {\cal O}_g/\partial g^{k-1}$ would allow
for the critical exponent that controls ${\cal O}_g$ to be immediately extracted from $\partial^{k-1} {\cal
E}/\partial g^{k-1}$.
This is not so, however. As an example, take a second order QPT $(k=2)$ with ${\cal O}_g = w_2$, where 
$\partial w_2/\partial u \sim |u-u_c|^{\rho-1} \rightarrow \infty$ as $g\rightarrow g_c = u_c$. By inspection of Eq.
(\ref{n:thOrder})
one then notes that the leading scaling of $\partial {\cal E}/\partial g$ will be governed by the same
exponent $\rho$
only if $m$ and $n$ are independent of $w_2$, or, depend on $w_2$ as a power with exponent $\ge 1$.
Whether this is the case typically requires that one has access to an exact solution of the model, and in
any event can only be determined on a case-to-case basis.
Turning to the logarithmic factors in (\ref{n:thOrder}) one realizes that these will cause logarithmic divergences
if one or several of the occupation parameters $w_0, w_{\uparrow}, w_{\downarrow}, w_2$ vanish at the transition
(cf. the parameterization in (\ref{OccupationParameters})). Such logarithmic corrections, multiplying the leading
scaling 
of $\partial^{k-1} {\cal E}/ \partial g^{k-1}$ inherited
from ${\cal O}_g$, thus signal a change of the dimension of the accessible local Hilbert space
as the system undergoes the transition. This is a useful and important property of the single-site
entanglement scaling not shared by the scaling of ${\cal O}_g$ or its derivatives.
One should here note that a spurious signaling of a $k^{\mbox{\small th}}$ order QPT by a divergence 
in $\partial^{k-1} {\cal E}/\partial g^{k-1}$
caused by a vanishing occupation parameter
is blocked by the constraint in the proposition that all lower-order derivatives of ${\cal E}$ are finite.
(Although maybe hard to realize, one may envision a system where one or several local basis states get excluded 
when tuning some parameter in the Hamiltonian [implying
the vanishing of an occupation parameter] without the occurrence of a QPT.) 

Using the diagnostics supplied by our proposition, are we guaranteed to catch {\em all} fermionic QPTs? The answer is
negative.
First, the diagnostics obviously fails for a QPT of {\em infinite order} \cite{ItoiMukaida}, a
Berezinski{\u \i}-Kosterlitz-Thouless (BKT) type transition being a case in point \cite{BKT}.
Secondly, and more insidious, a system may exhibit a QPT of finite order, but with
the single-site entanglement and its derivatives still remaining regular.    
This happens if all local basis states
$\ket{n}_j = \,\ket{0}_j, \, \ket{\,\uparrow}_j, \,\ket{\,\downarrow}_j$, and
$\ket{\,\uparrow \downarrow}_j$ become equally populated as one approaches the transition. As seen from
(\ref{n:thOrder}),
the $(k-1)^{\mbox{\small st}}$ derivative terms then vanish identically, killing the signal of the QPT. The simultaneous
vanishing
of $\partial {\cal E}/\partial g$ implies that ${\cal E}$ has a local extremum at the transition (expected to be a
maximum
since in this case all local basis states are equally represented in the make-up of the many-particle ground state).
However, one cannot {\em a priori} exclude that ${\cal E}$ is at an extremum without the occurrence of a QPT. Hence,
{\em an
extremum of the single-site entanglement does not necessarily signal a QPT}. Whether a QPT is present or
not in this case requires information beyond that provided by the entanglement measure.

Having exposed the general features of entanglement scaling at a fermionic QPT, let us 
look at two examples. 

{\bf Case studies.} \ Consider first the ordinary {\em 1D Hubbard model} 
\begin{equation}
H=-\sum_{i=1 \atop \sigma= \uparrow, \downarrow}^L(\hat{c}^{\dagger}_{i\sigma}\hat{c}_{i+1\sigma}+h.c.)+
u\sum_{i=1}^L\hat{n}_{i\uparrow}\hat{n}_{i\downarrow}
\end{equation}
with the first term describing hopping of electrons between neighboring sites, and with the second
term an effective on-site interaction of strength $u$. At half-filling of the lattice,
$n=1$,
the model exhibits a QPT at $u=0$, separating a Mott insulating phase $(u>0)$ from a metallic phase $(u<0)$.
The ground state energy density becomes non-analytic at the transition, but allows for an asymptotic 
power series expansion with 
all derivatives being finite and continuous \cite{MetznerVollhardt}. The QPT is thus of infinite order, and can
be shown to
belong to the BKT universality class \cite{Giamarchi}. 
As found by Gu {\em et al.}, the single-site entanglement
has a maximum at the transition. This reflects the equipartition of empty-, singly- and doubly occupied local states
when $u=0$ (non-interacting fermions). The transition is thus special on two
counts: it is of infinite order and it supports an equipartition of local states.
This makes it an exceptional example of a fermionic QPT, where no information 
can be deduced from the entanglement measure.

A metal-insulator transition can also be triggered when $u>0$ by connecting the system to a particle reservoir and
tuning
the chemical potential $g\equiv \mu$: When $n<1$ the system is metallic, but turns into an insulator at the critical
point $\mu_c = 2-4\int_0^{\infty} J_1(\omega)[\omega(1+\exp(\omega u /2))]^{-1}$ where $n=1$ \cite{LiebWu}. The
transition is second order with a
divergent charge susceptibility $\chi_c = \partial n/\partial\mu \sim |\mu - \mu_c|^{-1/2}$. As shown in Ref.
\onlinecite{Larsson}, the derivative of the critical single-site entanglement for finite $u$ is precisely given by
$\chi_c$, 
up to a multiplicative constant: $\partial {\cal E}/\partial\mu = -C(u) \chi_c$. In the limit $u\rightarrow \infty$
the empty local states get suppressed at the transition and the scaling of $\partial {\cal E}/\partial\mu$ picks up
a logarithmic correction \cite{Larsson}: $\partial {\cal E}/\partial\mu = \chi_c (\ln|\mu - \mu_c| +
\mbox{const.})/2\ln2$.
Both behaviors well illustrate our general discussion above: For finite $u$ the logarithms in Eq. (\ref{n:thOrder})
add up to the $u-$dependent constant $C(u)$, whereas in the limit $u\rightarrow \infty$ the entanglement measure
detects a change in the dimension of the local Hilbert space, as signaled by the logarithmic correction
to the leading scaling.

As a second example, let us consider the {\em 1D Hubbard model with long-range hopping}, introduced by
Gebhard and Ruckenstein \cite{GebhardRuckenstein}:
\begin{equation}
H=\sum_{\ell \neq
m=1 \atop \sigma = \uparrow, \downarrow}^L t_{\ell m}\hat{c}_{\ell \sigma}^{\dagger}\hat{c}_{m \sigma} 
+u\sum_{l=1}^L\hat{n}_{\ell \uparrow}
\hat{n}_{\ell \downarrow}
\end{equation}
with $t_{\ell m}=i(-1)^{(l-m)}(l-m)^{-1}$. The ground state energy density at half-filling is given by $e_0 =
(un-u_c(1-n)n)/4-(1/(24uu_c))[(u+u_c)^3-((u+u_c)^2-4uu_cn)^{3/2}]$ with $u_c=2\pi$ the critical point
\cite{GebhardRuckenstein}. This implies that
$w_2 = \partial e_0/\partial u$ has a discontinuity in its second order derivative with respect to $u$ at $u_c$
and hence the transition is third order.
From Eq. (\ref{SingleSiteEntanglement}) with $n=1$ it follows that the single site entanglement can be written as 
${\cal E}=-(1-2w_2)\log_2(1/2-w_2)-2w_2\log_2(w_2)$ when no magnetic field is present (i.e. $m=0$), and one
immediately verifies that $\partial^2{\cal E}/\partial u^2$ is also discontinuous at the transition point $u_c$.
Since the local basis states do {\em not} become equally populated at $u_c$ $-$ in contrast to the $u=0$
metal-insulator
transition of the ordinary Hubbard model $-$ the single-site entanglement here provides an accurate diagnostics of
the transition.

One can also drive a Mott-Hubbard metal-insulator transition by tuning the chemical potential when $u>u_c$,
in exact analogy with the ordinary Hubbard model. Expressing $n$ as a function of $\mu$, and applying the
Hellman-Feynman theorem to the ground state energy $e_0$ above, one obtains a discontinuity in  
$\partial n/\partial \mu$ at $\mu = \mu_c = \pi$ \cite{GebhardGirndt}. Eq. (\ref{n:thOrder}) immediately implies that $\partial
{\cal E}/
\partial \mu$ is also discontinuous at $\mu = \mu_c$, with the transition being second order.
In the limit $u \rightarrow \infty$ this discontinuity is multiplied by a logarithmic divergent factor
when $\mu \rightarrow \mu_{c-}$, reflecting
the suppression of empty states in this case. 

{\bf Summary.} \ We have shown that a generic finite-order quantum phase transition in a spin-1/2 fermionic lattice system
can be
consistently identified and characterized by studying the behavior of the single-site entanglement and its
derivatives with respect to the parameter that controls the transition. Extensions to cases where the transition is
driven by an interaction
or a field that couples to pairs or clusters of lattice sites (like the extended Hubbard model \cite{Anfossi1}) is
conceptually straightforward, albeit technically more demanding.
We hope to return to this problem in a future publication.

{\bf Acknowledgments.} \ We thank F. Gebhard and W. Metzner for valuable discussions.
D.L. thanks the Physics Department at Phillips Universit\"at Marburg for its hospitality.
H.J. acknowledges support from the Swedish Research Council under grant no. 621-2002-4947.


\end{document}